\documentclass[a4paper, onecolumn, preprint]{revtex4}
\usepackage{amsmath}
\usepackage{amsfonts}
\usepackage{graphicx}
\usepackage{color}
\usepackage{lscape}
\usepackage{amssymb}

\begin{document}
\title{Unified dynamic approach for simulating quantum tunneling and thermionic emission at metal/organic interface}

\author{Jiaqing Huang}
\affiliation{Department of Physics and State Key Laboratory of Luminescent Materials and Devices, South China University of Technology, Guangzhou 510640, China}

\author{Yijie Mo}
\affiliation{Department of Physics and State Key Laboratory of Luminescent Materials and Devices, South China University of Technology, Guangzhou 510640, China}

\author{Yao Yao}
\email{yaoyao2016@scut.edu.cn}
\affiliation{Department of Physics and State Key Laboratory of Luminescent Materials and Devices, South China University of Technology, Guangzhou 510640, China}

\begin{abstract}
Injection from metallic electrodes serves as a main channel of charge generation in organic semiconducting devices and the quantum effect is normally regarded to be essential. We develop a dynamic approach based upon the surface hopping (SH) algorithm and classical device modeling, by which both quantum tunneling and thermionic emission of charge carrier injection at metal/organic interfaces are concurrently investigated. The injected charges from metallic electrode are observed to quickly spread onto the organic molecules following by an accumulation close to the interface induced by the built-in electric field, exhibiting a transition from delocalization to localization. We compare the Ehrenfest dynamics on mean-field level and the SH algorithm by simulating the temperature dependence of charge injection dynamics, and it is found that the former one leads to an improper result that the injection efficiency decreases with increasing temperature at room-temperature regime while SH results are credible. The relationship between injected charges and the applied bias voltage suggests it is the quantum tunneling that dominates the low-threshold injection characteristics in molecular crystals, which is further supported by the calculation results of small entropy change during the injection processes. An optimum interfacial width for charge injection efficiency at the interface is also quantified and can be utilized to understand the role of interfacial buffer layer in practical devices.
\end{abstract}

\maketitle

\section{INTRODUCTION}

The successful industrial application of organic light emitting device (OLED) might be one of the great scientific achievements in the last decades. Researchers are still devoting effort to improve the efficiency so that it can be extended to more application scenarios such as illumination, and more comprehensive analysis of the microscopic working mechanism is thus highly demanded \cite{FHuang,JWang1,Lee,JWang2,JWang3}. Due to the cost control, the organic semiconductors are normally in amorphous phase and the contacts between conducting layer and electrodes are regarded to be poor. Moreover, different from the inorganic semiconductors in which intentional doping gives rise to numerous charge carriers, the main source of carriers in organic semiconductors stems from metallic electrodes. Therefore, the injection efficiency at the metal/organic interfaces has a significant impact on the overall performance of organic semiconducting devices.

In statistical physics, the charge injection at interfaces can be well described by the Richardson formula in the framework of thermionic emission theory \cite{Sze}. This theory is valid when the concentration of doped charges in inorganic semiconductors is sufficiently large so that the injected carriers would not drive the system far away from thermal equilibrium and the detailed balance principle matters. In this perspective, giving the structure of single-electron energy spectrum it is able to derive the injected current and the details of dynamical processes are not important. On the experimental side, therefore, the techniques such as the ultraviolet photoelectron spectroscopy are widely used to study the energy level alignment at these interfaces and thus the magnitude of the barrier for charge injection can be determined \cite{Koch,Fahlman1,Fahlman2,Fahlman3}. In contrast to the inorganic semiconductors, the charge injection at the metal/organic interfaces dominates the generation and interfacial recombination of charge carriers. Either thermionic emission or quantum tunneling may play the key role in the charge injection. Since the semiconductors are almost without free charges in the initial (as-prepared) stage, the injected charge easily enforces the system to the nonequilibrium state and the dynamical process of injection turns out to be essential. In addition, organic materials normally possess strong electron-phonon interactions, making the intrinsic mechanisms of charge injection more complicated \cite{Stokbro}.

In recent years, mixed quantum-classical dynamics (MQCD) has become an important tool to study the charge injection and transport properties in the organic semiconductors \cite{Troisi1,Troisi2,Wang1,Wang2,Wang3,Barbatti,Subotnik,Tully1,Wang4,Wang5,Wang6}. The MQCD can be generally performed in two different ways: The Ehrenfest dynamics and the surface hopping (SH) algorithm. The Ehrenfest dynamics based on the mean-field theory has a simple formulation in which the mean trajectory evolves on an effective potential energy surface (PES) for the electronic states and has been applied to the charge injection in the organic semiconductors by several groups \cite{Wu1,Wu2,Johansson,Xie,Junior}. By simulating the electron injection in a metal/polymer/metal structure, Wu and co-workers found that the injection of electron finally led to the formation of polaron or bipolaron and the electric field applied on the polymer could effectively reduce the interfacial potential barrier for charge injection from metallic electrode to polymer \cite{Wu1,Wu2}. Similarly, Johansson and Stafstr\"om simulated the formation of polaron in an isolated polymer chain \cite{Johansson}, while Fu and co-workers considered the similar issue in a metal/polymer structure \cite{Xie}. More recently, Ribeiro Junior and da Cunha obtained the polaron-type products through the careful investigations of the hole injection dynamics in polymers \cite{Junior}. In these works, the Ehrenfest dynamics were performed in the framework of the tight-binding Su-Schrieffer-Heeger model to describe the dynamic properties of the charge injection process. Although it works well for the simulation of elementary quasi-particles, it merely takes one effective PES into account and is not satisfactory when there is a large barrier and the quantum tunneling is remarkable. The discussion of charge injection in the practical organic semiconductors is still one of the pronounced topics especially when the device physicists want to understand the role of realistic device parameters such as the electric field, the interfacial morphology and so on.

It is known that due to the strong interaction between electrons and phonons, the organic systems may undergo non-adiabatic transitions among different PESs. The non-adiabatic phenomena are very common for a large number of situations, such as the atomic and molecular collision reactions, molecular photochemistry and photophysics, and photoexcitation of condensed phase systems \cite{Barbatti,Subotnik}. In recent years, the SH algorithm has become a crucial tool for the dynamic propagation of non-adiabatic systems in physics, chemistry and material sciences \cite{Barbatti,Subotnik,Tully1,Wang4,Wang5,Wang6}. In this algorithm, the nuclei move on an active PES with respect to the classical Newton equation and behave stochastic hops among different PESs based on the population flux among adiabatic electronic states. Tully's standard fewest switches surface hopping and its developed algorithms have been successfully used to study the charge transport in molecular crystals and the results always enlighten the importance of non-adiabatic transitions or quantum tunnelings for organic semiconductors \cite{Tully2,Wang7,Wang8,Wang9,Sifain,Wang10,Wang11,Wang12}. The present work is thus motivated to provide a unified approach based on the SH algorithm to simulate the charge injection process for different situations, no matter the quantum tunneling or thermionic emission dominates.

As typical organic semiconducting materials, the rubrene possesses large mobility and has lately attracted much research interest due to its low-threshold injection characteristics in heterojunction devices \cite{Pandey1,Pandey2,Xiang,Chen,He}. In this work, by embedding the classical Poisson's equation into the standard SH algorithm, the non-adiabatic dynamics of the charge injection at the metal/organic interfaces is investigated within the framework of a molecular crystal model. The generation of the injected charges will be firstly discussed. The injected charges calculated by the SH algorithm will be compared with the results by the Ehrenfest dynamics. The influence of the bias voltage applied at the metallic electrode on the injected charges and the optimum width of the interfacial layer for the charge injection will be investigated. Finally, the charge injection process at the metal/organic interfaces will be featured by the evolution of entropy change. The paper is organized in the following sequence. The model for the metal/organic system and the dynamic evolution method are given in Section \uppercase\expandafter{\romannumeral2}. The results and discussion are presented in Section \uppercase\expandafter{\romannumeral3}. Finally, in Section \uppercase\expandafter{\romannumeral4}, the main conclusions are drawn.

\section{MODEL AND METHOD}

Throughout this work, we take rubrene as instance to show the simulation results from our developed dynamic approach. Without loss of generality, a metallic electrode and the rubrene molecules are placed head-to-tail to construct a one-dimensional metal/organic structure, in which totally 160 sites are considered and we take 70 sites on the left to mimic the atoms of metallic electrode and the other 90 sites on the right for the organic molecules, respectively. Moreover, as a crucial consideration of this work, we notice that the morphology of the interface is essential for the charge injection. Specifically, as the organic materials are normally soft and amorphous, there is remarkable permeability of organic molecules into the metallic electrode. Therefore, a certain number of sites in the electrode close to the organic side will be regarded as the interfacial layer.

The model Hamiltonian consists of three terms as follows:
\begin{equation}\label{total hamiltonian}
H=H_{\rm ele}+H_{\rm lat}+H_{\rm ext}.
\end{equation}
The first term in Eq.~(\ref{total hamiltonian}) is the electronic Hamiltonian expressed as
\begin{equation}\label{electronic hamiltonian}
H_{\rm ele}=-\sum_{n}t_{n} (\hat{c}_{n+1}^{\dagger} \hat{c}_{n}+{\rm h.c.}),
\end{equation}
where $\hat{c}_{n}^{\dagger}$ ($\hat{c}_{n}$) creates (annihilates) an electron on $n$-th site and $t_n=t_{0(1)}-{\alpha}_n (u_{n+1}-u_n)$ is the nearest-neighbor hopping integral in which the vibronic coupling is involved, namely, with ${\alpha}_n$ being the vibronic coupling strength and $u_n$ being the displacement of $n$-th site. Herein, $t_0$ is the hopping constant in both electrode and organic material, and $t_1$ is the coupling between them. The vibronic coupling ${\alpha}_n=0$ for metallic sites far away from interface so that it is gapless. For the organic molecules, we take uniform vibronic couplings, namely ${\alpha}_n={\alpha}$, so that the initial state of the organic molecules is dimerized due to the Peierls instability and the gap is determined by the value of ${\alpha}$. In this situation, there naturally exists a barrier between electrode and organic material. In the interfacial layer, the vibronic coupling strength in the side of metallic electrode is given by a half Gaussian function, namely,
\begin{equation}\label{vibronic coupling}
{\alpha}_n={\alpha} \exp\left[-4(n-n_0)^2/ W^2\right],
\end{equation}
with $n_0$ being the last site of the metallic electrode and $W$ being the width of the interfacial layer. Here, $W$ is an essential parameter in our model to characterize the interface.

The second term in Eq.~(\ref{total hamiltonian}) represents the elastic potential and kinetic energy of the molecules,
\begin{equation}\label{lattice hamiltonian}
H_{\rm lat}=\frac{K}{2}\sum_{n} (u_{n+1}-u_{n})^2+\frac{M}{2}\sum_{n} \dot{u}_{n}^2,
\end{equation}
with $K$ being the elastic constant and $M$ being the mass of molecules.

The third term of Eq.~(\ref{total hamiltonian}) describes the contribution from the external field and has the following form
\begin{equation}\label{potential energy hamiltonian}
H_{\rm ext}=\sum_{n} U_{n}(t) \hat{c}_{n}^{\dagger}\hat{c}_{n},
\end{equation}
where $U_{n}(t)$ is the potential energy induced by the applied bias voltage and the electric field. In the metallic electrode a bias voltage is applied and written as $U_{n}(t)=\vert e \vert V(t)$. In the organic molecules an electric field $E_{n}(t)$ along the $-\hat{x}$ direction is thus generated and then the potential energy is determined by $U_{n}(t)=-\vert e \vert V_{n}(t)$. As another important input of our approach, we consider to embed the device modeling into the dynamic simulations. That is, considering the feedback of the distribution of the injected charges onto the electric field, $E_{n}(t)$ and $V_{n}(t)$ are given by the Poisson's equation \cite{Pflumm,Christ,Li}:
\begin{equation}\label{electric field}
E_{n}(t)=E_{n-1}(t)-{\frac{\vert e \vert a_{0}{\rho}_{n}(t)}{{\varepsilon}_{\rm r} {\varepsilon}_0 V_{\rm rub}}},~~~V_{n}(t)=V_{n-1}(t)-E_{n}(t) a_{0},
\end{equation}
where $e$ is the elementary charge of an electron, $\varepsilon_{\rm r}$ is the relative dielectric constant, $\varepsilon_0$ is the permittivity of vacuum, $a_0$ is the lattice constant, $V_{\rm rub}$ is the occupying volume of a single molecule and ${\rho}_n$ is the injected charges on the $n$-th site. In the practical simulations, in order to minimize the numerical errors, we turned on the external field smoothly by using a half Gaussian function, that is, $V(t)=V_0 \exp\left[-(t-t_{\rm c})^2/ {t_{\rm w}^2}\right]$ for $0 \textless{t}\textless{t_{\rm c}}$ and $V(t)=V_0$ for $t \geq{t_{\rm c}}$ with $t_{\rm c}$ being a smooth turn-on period and $t_{\rm w}$ the width. We set $t_{\rm c}=30~{\rm fs}$ and $t_{\rm w}=25~{\rm fs}$ which are optimum values for numerical accuracy through our testing. The electric field $E_1(t)$ at the first organic molecule is also turned on smoothly in the same way, and the initial value $E_{1}$ is determined by $V_0=La_0 E_{1}$ with $L$ being the site number of the organic molecules.

The model parameters in this work are obtained from the earlier work by Troisi for investigating the charge transport of rubrene, namely, $t_0=1150~{\rm cm^{-1}}$, ${\alpha}=3980~{\rm cm^{-1}/\r{A}}$, $K=48200~{\rm amu/ps^2}$, $M=532~{\rm amu}$ \cite{Troisi2}, and $a_0=3.4~{\rm \r{A}}$, $V_{\rm rub}=760~{\rm {\r{A}}^3}$ by our computations. For most organic materials, the dielectric constant normally ranges between 3 and 4 \cite{Arkhipov}, and we set it to 4 in this work. For description of the metal/organic interface with a potential barrier, we set $t_1=0.8t_0$.

The motion of the sites can be described by the Newtonian equation. As we want to concurrently take the thermionic emission into account, the temperature should also be involved as a parameter in our model. Hence, the Langevin equation is employed as follows:
\begin{equation}\label{force}
F_{n} (t)=M\ddot{u}_{n}=-K[2u_{n}-u_{n+1}-u_{n-1}]+\alpha_n[\rho_{n,n+1}-\rho_{n-1,n}+\rho_{n+1,n}-\rho_{n,n-1}]-\gamma M\dot{u}+\xi_n.
\end{equation}
Herein, $F_{n} (t)$ represents the force exerted on the ${n}$-th site and the density matrix $\rho$ is given by $\rho_{n,{n'}}=\sum_{\mu}\psi_{\mu,n}^{\ast} (t)f_{\mu} \psi_{\mu,n'} (t)$ with $f_{\mu}$ (equals to 0, 1 or 2) being the time-independent distribution function determined by the initial occupation of electrons, $\gamma$ is the friction coefficient which is set to $0.01~{\rm ps^{-1}}$, and $\xi_n$ is a Markovian Gaussian random force with standard deviation $(2\gamma Mk_{\rm B}T/{\Delta}t)^{1/2}$ to mimic the thermal effect. With these considerations, the lattice displacement $u_{n} (t_{j+1})$ and the velocity $\dot{u}_{n} (t_{j+1})$ are respectively given by
\begin{equation}\label{lattice displacement}
u_{n}(t_{j+1})=u_{n} (t_j)+\dot{u}_{n}(t_j){\Delta}t,~~~\dot{u}_{n}(t_{j+1})=\dot{u}_{n}(t_j)+{\frac{F_{n} (t_j)}{M}} {\Delta}t.
\end{equation}

The time evolution of the electronic wavefunction is given by the time-dependent Schr\"odinger equation:
\begin{equation}\label{S equation}
i\hbar\frac{\partial{\psi_{\mu,n}(t)}}{\partial{t}}=-t_{n} \psi_{\mu,n+1} (t)-t_{n-1} \psi_{\mu,n-1} (t),
\end{equation}
where $\psi_{\mu,n}$ is the $\mu$-th eigenstate of the Hamiltonian (\ref{total hamiltonian}) on the $n$-th site. Herein, in order to numerically solve the electronic wavefunction, the integration time step ${\Delta}t$ in the discretization procedure must be set to be sufficiently small, namely below the order of the bare phonon frequency $\omega_{\rm Q}=\sqrt{4K/M}$. It is thus set to be 0.2 {\rm fs} except for the trivial crossings. Therefore, the solution of the time-dependent Schr\"odinger equation is finally written as \cite{Huang}
\begin{equation}\label{solution}
\psi_{\mu}(t_{j+1})=\sum_{\nu} \left\langle\varphi_{\nu}\mid{\psi_{\mu} (t_{j})}\right\rangle \exp\left[-i\varepsilon_{\nu} {\Delta}t/{\hbar}\right] \varphi_{\nu},
\end{equation}
where $\varphi_{\nu}$ and $\varepsilon_{\nu}$ are the instantaneous eigenfunctions and eigenvalues of the Hamiltonian, respectively.

The eigenfunctions of the Hamiltonian can be expressed with the original diabatic orbitals, $\varphi_{\nu}=\sum_n p_{n,\nu} \mid n\rangle$. The electronic wavefunction is described with a linear expansion of these eigenfunctions, $\psi=\sum_{\nu} c_{\nu}\varphi_{\nu}$. The time-dependent Schr\"odinger equation is then written as
\begin{equation}\label{rewritten S equation}
\dot{c}_{\nu}={\frac{1}{i\hbar}}c_{\nu} \varepsilon_{\nu}-\sum_{\mu\not=\nu} c_\mu \sum_n \dot{u}_n d_{\nu,\mu}^n,
\end{equation}
with $d_{\nu,\mu}^n=\left\langle\varphi_{\nu}\mid {\rm d}\varphi_{\mu}/{\rm d}u_n\right\rangle$ being the non-adiabatic coupling which is obtained by the Hellmann-Feynman theorem, i.e.,
\begin{equation}\label{non-adiabatic coupling}
d_{\nu,\mu}^n=\frac{\alpha_n [p_{n,\nu} (p_{n-1,\mu}-p_{n+1,\mu})+p_{n,\mu} (p_{n-1,\nu}-p_{n+1,\nu})]}{\varepsilon_{\mu}-\varepsilon_{\nu}}.
\end{equation}

If the quantumness of the system is strong enough, the charge injection will be dominated by the quantum tunneling which is featured by the non-adiabatic transitions among different PESs adhere to the electrode and organic molecules, respectively. In inorganic semiconductors, one can simply consider a scattering process through, for example, a triangle-shaped energy barrier to simulate this tunneling effect. In organic semiconductors, however, the electron is self-trapped and relatively localized such that it is not easy for the electron to hop across the barriers by their own kinetic energy, and the electron-phonon interactions normally serve as the main driving force to induce the quantum tunnelings. To this end, we have to adopt the Tully's standard SH algorithm to take this tunneling effect into the simulations, and the hopping probability from the active surface $\nu$ to another surface $\mu$ is given by \cite{Tully2}
\begin{equation}\label{hopping probability}
g_{\nu,\mu}={\Delta}t\cdot {\frac{2{\rm Re}(c_{\nu} c_{\mu}^{\ast}\sum_n \dot{u}_n d_{\nu,\mu}^n)}{c_{\nu}^{\ast}c_{\nu}}}.
\end{equation}
If $g_{\nu,\mu}\textless{0}$, it is reset to be zero. A uniform random number $\zeta$ is generated and then a surface hop takes place if $\sum_{X\not=\nu}^{\mu-1} g_{\nu,X}\textless{\zeta}\leq{\sum_{X\not=\nu}^{\mu}g_{\nu,X}}$. A velocity adjustment is then made to conserve the total energy if ${\varepsilon_{\mu}-\varepsilon_{\nu}}\textless{\frac{M}{2} \sum_n \dot{u}_n^2}$, i.e.,
\begin{equation}\label{velocity adjustment}
\dot{u}'_n=\dot{u}_n+d_{\nu,\mu}^n \frac{A}{B} \left[\sqrt{1+2(\varepsilon_{\nu}-\varepsilon_{\mu})\frac{B}{A^2}}-1\right],
\end{equation}
where $A=\sum_n M\dot{u}_n d_{\nu,\mu}^n$ and $B=\sum_n Md_{\nu,\mu}^n\cdot d_{\nu,\mu}^n$. It is noted that the hop cannot occur if $\varepsilon_{\nu}\textless{\varepsilon_{\mu}-A^2/2B}$.

It is known that the trivial crossings are very common in large systems involving many electronic states due to the rapid change with time of the non-adiabatic couplings \cite{Granucci,Fernandez-Alberti}. Adaptive time intervals \cite{Wang7,Fernandez-Alberti,Virshup,Sporkel} and manual surface hops \cite{Fernandez-Alberti,Fabiano} have been proven to be valid in SH simulations to deal with the trivial crossings. Considering the computational cost and the size of our model system, we first adjust ${\Delta}t$ to be 0.02 {\rm fs} for the calculation of the explicit hopping probabilities and then make manual hops if ${\Delta}t$ is not small enough. For simplicity, we only consider the electron residing on top of Fermi surface that stochastically hops from the active PES to another in a time step. To obtain smooth averaged curves of the time-dependent injected charges, 100 realizations for each simulation are performed.

\section{RESULTS AND DISCUSSION}

\begin{figure}
\centering
\includegraphics[scale=1.7]{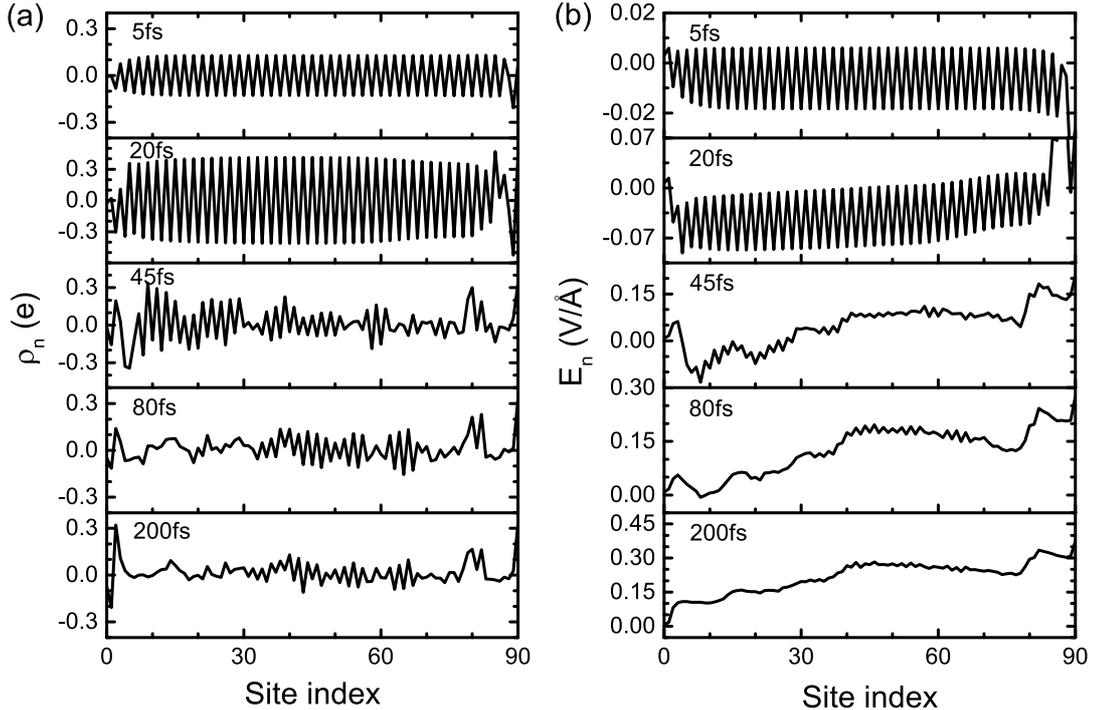}
\caption{(a) The injected charge distribution ${\rho}_n(t)$ and (b) the corresponding electric field distribution $E_n(t)$ in the organic molecules at different time points with $W=20a_0$, $V_0=5~{\rm V}$ and $T=300~{\rm K}$.
}\label{figure1}
\end{figure}

We first discuss the dynamical process of the injected charges in the organic molecules. By using the SH algorithm, we calculate the injected charge distribution ${\rho}_n(t)$ and the corresponding electric field distribution $E_n(t)$ in the organic molecules at different time points with interfacial width $W=20a_0$, bias voltage $V_0=5 {\rm V}$ and temperature $T=300 {\rm K}$. As shown in Fig.~\ref{figure1}(a), driven by the applied bias voltage, charges initially residing in the metallic electrode are injected into the organic molecules by overcoming the interfacial barrier induced by the Peierls instability and then quickly spread onto the whole organic molecules. Afterward, enforced by the built-in electric field $E_n(t)$, the injected charges are redistributed and accumulated on the left hand side of the organic molecules near the metal/organic interface after a relaxation time of $\sim80~{\rm fs}$. This is because we do not consider the other electrode in the devices such that there is not a drain for the carriers; we are dealing with a single contact for the first step. From Fig.~\ref{figure1}(b), one can find that $E_n(t)$ becomes larger with more charges being injected into the organic layer, which in turn further promotes the redistribution of the injected charges. It is known that the formation of stable localized polaron generally takes place in the charge injection process in the $\pi-$conjugated systems such as polyacetylene \cite{Wu1,Johansson,Xie,Rakhmanova}. However, it is also reported that the transport characteristics of molecular crystals, such as pentacene and rubrene, generally show high mobility and delocalized quantum carriers \cite{Podzorov,Ostroverkhova,Minari,Gershenson}, which is quite different from that in the $\pi-$conjugated systems. Benefitting from the utilization of quantum dynamic approach combined with the Poisson's equation, the phenomenon of the delocalization-localization transition of the injected charges in organic molecular materials under the applied bias voltage can be well simulated.

\begin{figure}
\centering
\includegraphics[scale=1.3]{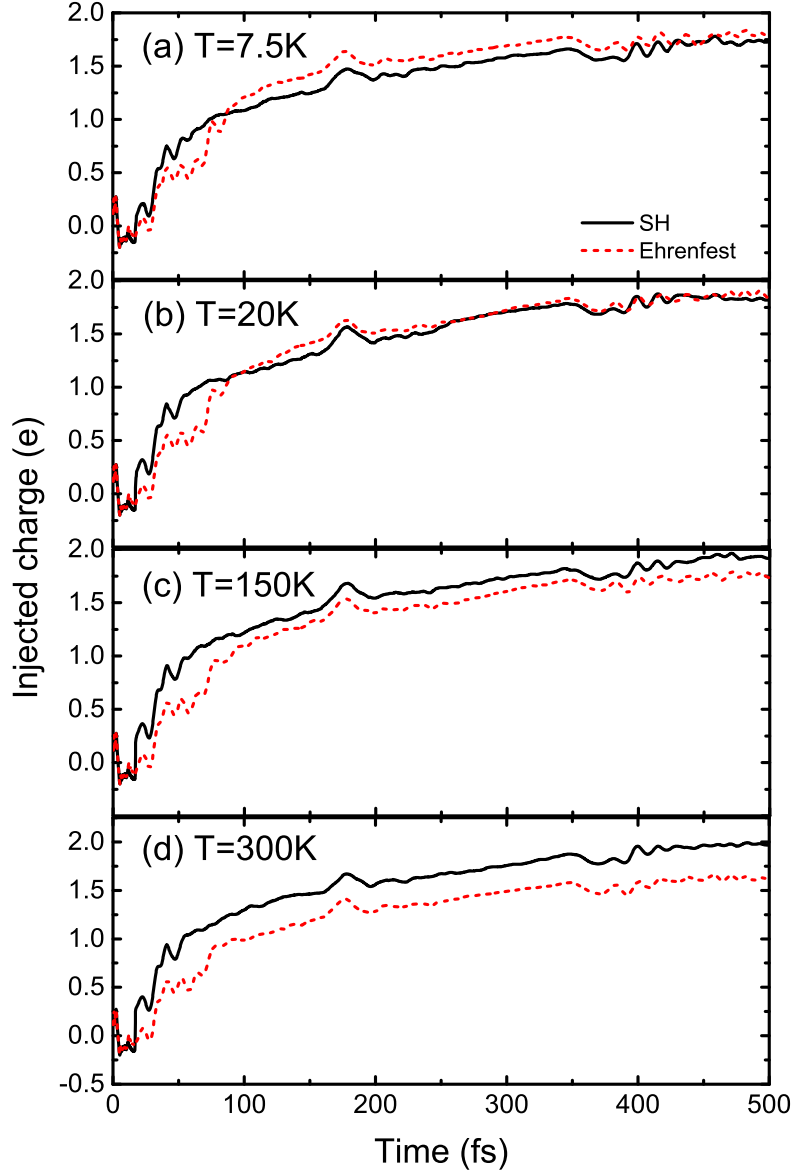}
\caption{Time evolution of the injected charge into the organic molecules for various temperatures $T=$ (a) $7.5~{\rm K}$, (b) $20~{\rm K}$, (c) $150~{\rm K}$ and (d) $300~{\rm K}$ for the comparison between the SH algorithm and the Ehrenfest dynamics. The other parameters are: $W=20a_0$ and $V_0=5~{\rm V}$.
}\label{figure2}
\end{figure}

In order to show the crucial role of quantum tunneling in charge injection, we give a comparison of efficiency for the charge injection at the metal/organic interfaces by individually using the SH algorithm and the Ehrenfest dynamic approach. Fig.~\ref{figure2} displays the time evolution of the injected charges $Q$ at four typical temperatures from low to high. The other parameters are $W=20a_0$ and $V_0=5~{\rm V}$. One can observe that $Q$ increases gradually with evolution time and finally converges to different values at several hundreds of femtoseconds. The noticeable differences of the dynamic properties should be emphasized for these two methods. It can be found from Fig.~\ref{figure2}(a) that, $Q$ calculated by the Ehrenfest dynamics is larger than that by the SH algorithm at ultralow temperature $T=7.5~{\rm K}$. However, with temperature increasing, $Q$ calculated by the SH algorithm becomes close to that of the Ehrenfest dynamics at $T=20~{\rm K}$ (see Fig.~\ref{figure2}(b)) and then gets to be larger at $T=150~{\rm K}$ and $T=300~{\rm K}$ (see Fig.~\ref{figure2}(c) and (d), respectively) due to the quantum tunneling represented by the non-adiabatic transition between PESs adhere to the electrode and organic molecules.

\begin{figure}
\centering
\includegraphics[scale=1.3]{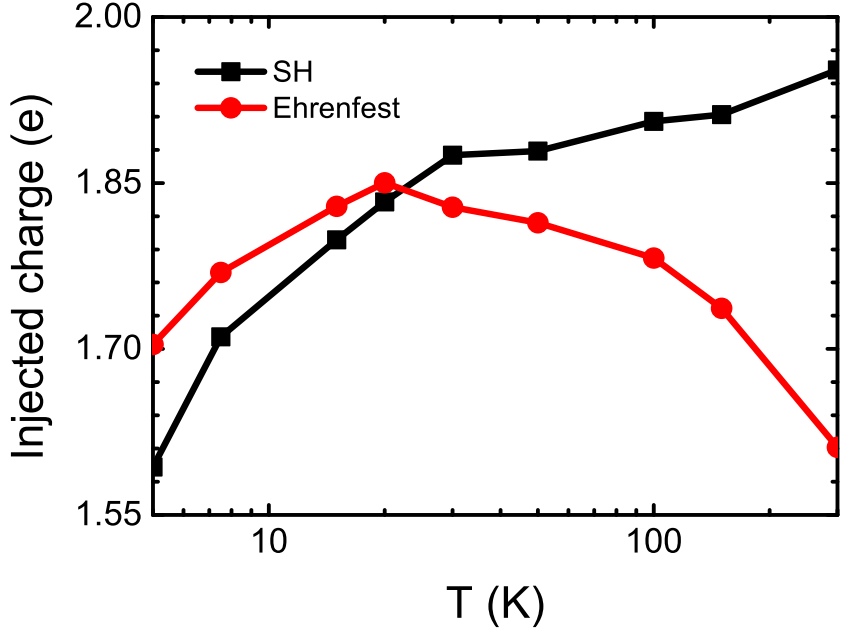}
\caption{Injected charge versus the temperature for the comparison between the SH algorithm and the Ehrenfest dynamics with $W=20a_0$ and $V_0=5~{\rm V}$.
}\label{figure3}
\end{figure}

To further manifest the advantages of the SH algorithm for simulating the charge injection, we calculate the mean value of injected charges $Q$ over the smooth time range between $400~{\rm fs}$ and $500~{\rm fs}$ for more typical temperatures. Fig.~\ref{figure3} shows the temperature dependence of $Q$ obtained by the SH algorithm and the Ehrenfest dynamics with $W=20a_0$ and $V_0=5~{\rm V}$. The curves can be analyzed in two different temperature regimes. One is from $5~{\rm K}$ to $30~{\rm K}$, in which $Q$ obtained by the Ehrenfest dynamics is slightly larger than that from the SH algorithm, and the other is from $30~{\rm K}$ to $300~{\rm K}$, in which $Q$ obtained by the SH algorithm is significantly larger. More importantly, the temperature dependence of injected charges calculated from Ehrenfest dynamics shows a counterintuitive result that the charge injection efficiency decreases with increasing temperature, so we can conclude the Ehrenfest dynamics breaks down at room temperature. The reason is that, at low temperature the non-adiabatic transitions are largely suppressed so the mean-field treatment of PES in Ehrenfest dynamics shares the similar results with the SH algorithm. At room temperature, on the other hand, non-adiabatic transitions obviously dominate the charge injection so we have to use SH algorithm to properly simulate this process. It is also found that the lineshape deviates from that of Richardson formula, namely $\sim T^2\exp(-\phi/k_{\rm B}T)$ with $\phi$ being the injection barrier. Our result exhibits larger injection efficiency than that from the Richardson formula within $30~{\rm K}$ to $300~{\rm K}$, which is explicitly the consequence of involving quantum effects.

\begin{figure}
\centering
\includegraphics[scale=1.3]{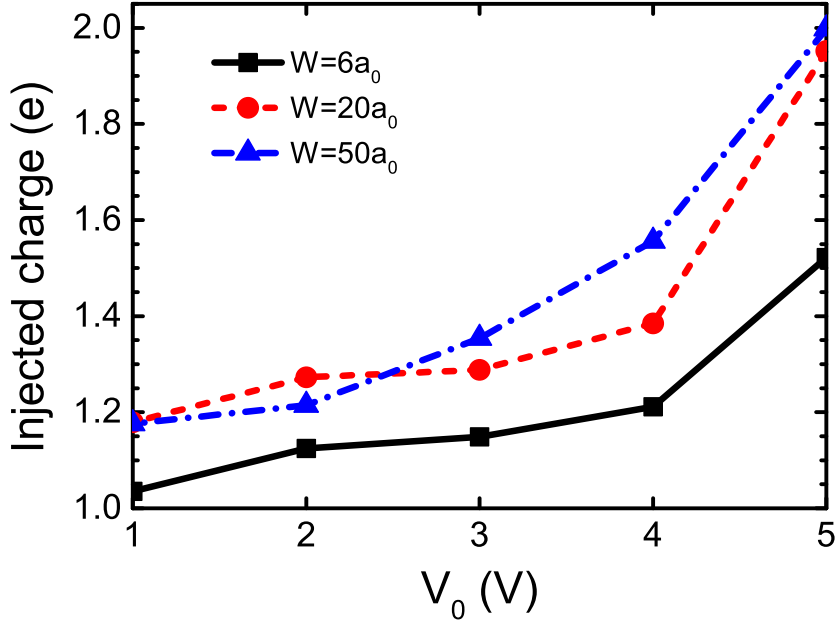}
\caption{Injected charge versus the applied bias voltage $V_0$ by using the SH algorithm for various interfacial width $W$ with $T=300~{\rm K}$.
}\label{figure4}
\end{figure}

For the sake of device modeling, our dynamic approach has to be working with the influence of the bias voltage $V_0$ applied at the metallic electrode on the charge injection efficiency. To this end, three different values of $W$ are chosen for the calculation of the total injected charges at $T=300~{\rm K}$, which is displayed in Fig.~\ref{figure4}. It shows a monotonic increase of $Q$ with the increase of $V_0$ for three values of $W$. When a larger $V_0$ is applied at the metallic electrode, the initial electric field in the organic layer becomes larger, which has a stronger promoting effect to the charge injection. Surprisingly, when the value of $V_0$ is low to be $1~{\rm V}$ which is about half of the bandgap of rubrene ($\sim 2.2~{\rm eV}$), the charges can still be injected into the organic molecules efficiently. In spite of the combined electric field simulation in our model, this result shows the low-threshold injection characteristics of these molecular crystals such as rubrene, which is quite in agreement with the experimental results \cite{Pandey1,Pandey2,Xiang,Chen,He}. Moreover, from Fig.~\ref{figure4} one can also find that $Q$ becomes larger when the larger $W$ is applied in the range of high $V_0$ while the trend is not obvious when $V_0$ is lower than $3~{\rm V}$, implying the efficiency of charge injection at the metal/organic interfaces is closely related to the width of the interfacial layer which should be further investigated.

\begin{figure}
\centering
\includegraphics[scale=1.3]{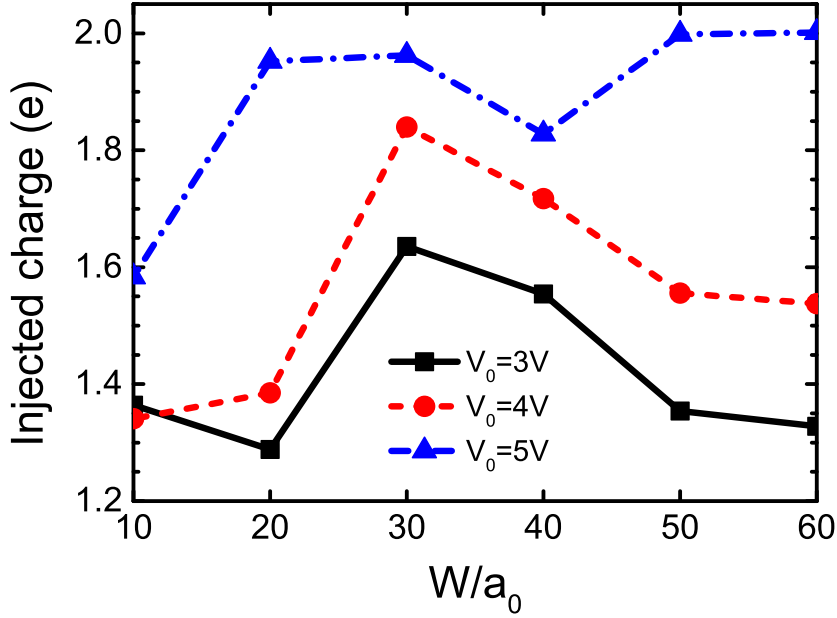}
\caption{Injected charge versus the interfacial width $W$ by using the SH algorithm for various bias voltage $V_0$ with $T=300~{\rm K}$.
}\label{figure5}
\end{figure}

As well known, interfaces are essential for organic semiconductors in any way. In our previous work, we have discussed an optimum width of donor/acceptor (D/A) interface for the dissociation of the charge-transfer (CT) state \cite{Huang}. In the practical fabrication of devices, the insertion of a buffer layer between the metallic electrode and the organic layer can effectively facilitate the injection of charge \cite{Hou,Guo1,Guo2,Guo3} and Hou's group has obtained an optimal buffer layer width of several nanometers based on the model calculations and the experimental investigations \cite{Hou}. Here with our dynamic approach, we can also determine an optimum width of the metal/organic interface for the charge injection. Fig.~\ref{figure5} displays the injected charges $Q$ among various interfacial widths at certain $V_0$ with $T=300~{\rm K}$. For the cases of $V_0=3~{\rm V}$ and $V_0=4~{\rm V}$, with the increase of the interfacial width, $Q$ dramatically increases for $W<30a_0$ followed by a remarkable decrease. Differently, for the case of $V_0=5~{\rm V}$, $Q$ first dramatically increases and then behaves a slight increase except for the case $W=40a_0$. As a consequence here, the optimum value of the interfacial width for the charge injection is determined to be $30a_0$ ($\sim10~{\rm nm}$) in our case, which is consistent with Hou's results \cite{Hou}.

\begin{figure}
\centering
\includegraphics[scale=1.3]{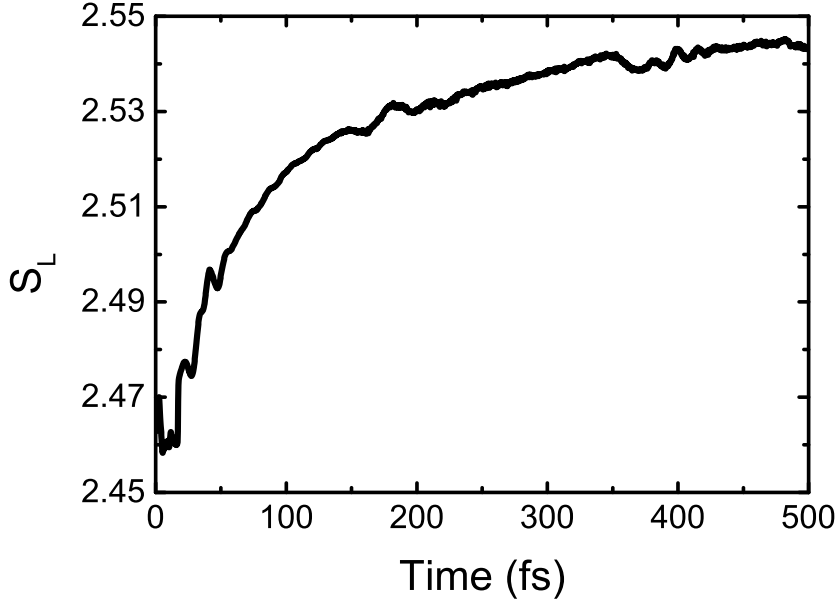}
\caption{Time evolution of the von Neumann entropy $S_{\rm L}$ of the injected charges with $W=20a_0$, $V_0=5~{\rm V}$ and $T=300~{\rm K}$.
}\label{figure6}
\end{figure}

Before ending this section, we would like to discuss more on the quantumness of the charge injection process at the metal/organic interfaces. As stated above, the classical thermionic emission theory can not be directly applied to the organic semiconductors; lots of quantum effects have been discussed in this systems especially taking photoemission into consideration \cite{Barbatti,Subotnik,Nelson}. To this end, we calculate the von Neumann entropy of the injected charges, which is defined as \cite{Lauchli,Amico}:
\begin{equation}\label{von Neumann entropy}
S_{\rm L}=-{\rm Tr}(\rho_{\rm L} \ln{\rho}_{\rm L}),
\end{equation}
where $\rho_{\rm L}$ is the reduced density matrix of the charges in organic molecules. Fig.~\ref{figure6} exhibits the time evolution of the von Neumann entropy $S_{\rm L}$ of the system in the charge injection process with $W=20a_0$, $V_0=5~{\rm V}$ and $T=300~{\rm K}$. It is found that $S_{\rm L}$ increases gradually with evolution time and finally saturates to a constant. We notice that, the total increase of entropy is roughly by $3\%$, not significant compared with our expectation as we expected that following more charges are injected the entropy would largely increases making the quantumness quickly lost. In our previous work, we have also present a coherent scenario of the dissociation of the CT state at the D/A interfaces in the organic photovoltaics \cite{Huang}. Combined these two studies, our findings may enlighten the understanding of the quantumness of charge dynamics in organic semiconductors which may facilitate the establishment of a systematic device modeling method specifically for this system.

\section{CONCLUSION AND OUTLOOK}

In summary, we develop a unified dynamic approach based upon surface hopping algorithm combined with Poisson's equation to investigate the charge injection process at metal/organic interface. It is found that the charges can be dynamically injected into the organic materials and then quickly spread onto the molecules in a delocalized fashion followed by an accumulation induced by the built-in electric field. The injected charges obtained by the SH algorithm are compared with that by the Ehrenfest dynamics, and the former one works well for simulating efficient charge injection at room temperature while the latter one is merely valid at ultralow temperature. The influence of the bias voltage applied at the metallic electrode on the charge injection efficiency is discussed and the injected charges behave a significant increase by increasing applied bias voltage. With the assistance of quantum effect, the charges could be injected into the organic molecules efficiently even if the bias voltage is small compared with the bandgap of the organic material, which can be utilized to explain the low-threshold feature. The width of $\sim10~{\rm nm}$ is regarded to be the optimum value of the interfacial layer for the charge injection at the metal/organic interfaces. In addition, the von Neumann entropy provides a further insight into the quantumness of the charge injection dynamics.

In inorganic semiconductors the thermal equilibrium is basically valid in an approximate manner, since the thermal motions of atoms are almost completely disordered. In contrast, the molecular vibrations in organic materials are always concentrated at some specific modes so that the quantum coherence of electrons could be well protected to keep quantumness alive. In organic photovoltaic devices, the coherent exciton dynamics has been demonstrated in many experiments and widely discussed \cite{Lienau1,Scholes1,Scholes2,Lienau2}. As the majority charge carriers in organic semiconductors stem from injection at the interfaces other than doping, the normal thermionic emission theory is not sufficient for us to understand the conductivity in these devices. A dynamic scenario serves as an alternative perspective. With this unified dynamic approach in hand, we can in the future establish a systematic device modeling method to mimic the device performance giving the material parameters obtained from quantum chemistry computations. We believe this first-principle framework could help us simulate the organic semiconducting devices in a more appropriate and efficient way.

\section*{ACKNOWLEDGMENTS}

The authors gratefully acknowledge support from the National Natural Science Foundation of China (Grant No.~91833305 and 11974118), the Key R$\&$D Project of Guangdong Province (Grant No.~2020B0303300001) and the Fundamental Research Funds for the Central Universities (Grant No.~2019ZD51).

\end{document}